\documentclass[runningheads]{llncs}
\usepackage[nolist]{acronym}

\usepackage{cite}
\usepackage{colortbl}
\usepackage{hyperref}

\usepackage{graphicx}
\usepackage{enumitem}
\usepackage{array}
\usepackage{diagbox}
\usepackage{colortbl}
\usepackage{hhline}
\usepackage{graphicx}
\usepackage[british]{babel} 
\usepackage{siunitx}
\usepackage{subcaption}
\usepackage{array}
\newcolumntype{R}[1]{>{\raggedleft\arraybackslash}p{#1}}
\usepackage{multicol}
\usepackage{xcolor}
\definecolor{webScarping}{HTML}{468B97}
\definecolor{performanceTest}{HTML}{F3AA60}

\usepackage[breakable, hooks]{tcolorbox}
\usepackage{tikz}

\usepackage{bbding}

\begin{document}
\begin{acronym}
    \acro{CPU}{Central Processing Unit}
    \acro{LAN}{Local Area Network}
    \acro{CSS}{Cascading Style Sheet}
    \acro{PNG}{Portable Network Graphics}
    \acro{JPEG}{Joint Photographic (Expert) Group format}
    \acro{WEBP}{Web Picture format}
    \acro{SVG}{Scalable Vector Graphics}
    \acro{GIF}{Graphics Interchange Format}
    \acro{AVIF}{AV1 Image File Format}
    \acro{APNG}{Animated Portable Network Graphics}
    \acro{BMP}{Bitmap file}
    \acro{ICO}{Microsoft Icon}
    \acro{TIFF}{Tagged Image File Format}
    \acro{MPO}{Multi Picture Object}
    \acro{PSD}{Photoshop Document}
    \acro{BPP}{Bits per Pixel}
    \acro{FP}{First Paint}
    \acro{FCP}{First Contentful Paint}
    \acro{TTFB}{Time To First Byte}
    \acro{LCP}{Largest Contentful Paint}
    \acro{TBT}{Total Blocking Time}
    \acro{PLT}{Page Load Time}
    \acro{SSIM}{Structural Index Similarity}
    \acro{ANOVA}{Analysis of Variance}
    \acro{BPP}{Bits per Pixel}
\end{acronym}
\title{Web Image Formats: Assessment of Their Real-World-Usage and Performance across Popular Web Browsers}

\titlerunning{Comparison of Web Image Formats}

\author{Benedikt Dornauer\inst{1,2}\orcidID{0000-0002-7713-4686} \and
Michael Felderer\inst{1,2,3}\orcidID{0000-0003-3818-4442} }
\authorrunning{Dornauer et al.}

\institute{University of Innsbruck, 6020 Innsbruck, Austria \\ \email{benedikt.dornauer@uibk.ac.at}\and
University of Cologne, 50923 Cologne, Germany \and
German Aerospace Center (DLR), Institute for Software Technology, 51147 Cologne, Germany\\ 
\email{michael.felderer@dlr.de}}

\maketitle              
\begin{abstract}
In 2023, images on the web make up 41 \%  of transmitted data, significantly impacting the performance of web apps. Fortunately, image formats like WEBP and AVIF could offer advanced compression and faster page loading but may face performance disparities across browsers. Therefore, we conducted performance evaluations on five major browsers - Chrome, Edge, Safari, Opera, and Firefox - while comparing four image formats. The results indicate that the newer formats exhibited notable performance enhancements across all browsers, leading to shorter loading times. Compared to the compressed JPEG format, WEBP and AVIF improved the Page Load Time by 21\% and 15\%, respectively. However, web scraping revealed that JPEG and PNG still dominate web image choices, with WEBP at 4\% as the most used new format. Through the web scraping and web performance evaluation, this research serves to (1) explore image format preferences in web applications and analyze distribution and characteristics across frequently-visited sites in 2023 and (2) assess the performance impact of distinct web image formats on application load times across popular web browsers.
\end{abstract}
\section{Introduction}
\label{sec:introduction}
Images are one of the critical elements that can make a web application more appealing and competitive, as they can attract and engage users with the content. The rationale behind this is simple: Users tend to focus on images before they read the textual content, which makes them a powerful tool for capturing user interest \cite{wangVisualDesignWeb2021}. Moreover, images can illustrate stories, are used to display ads, or simply represent products or services. 

\begin{figure}
    \centering
    \includegraphics[width=0.9\textwidth]{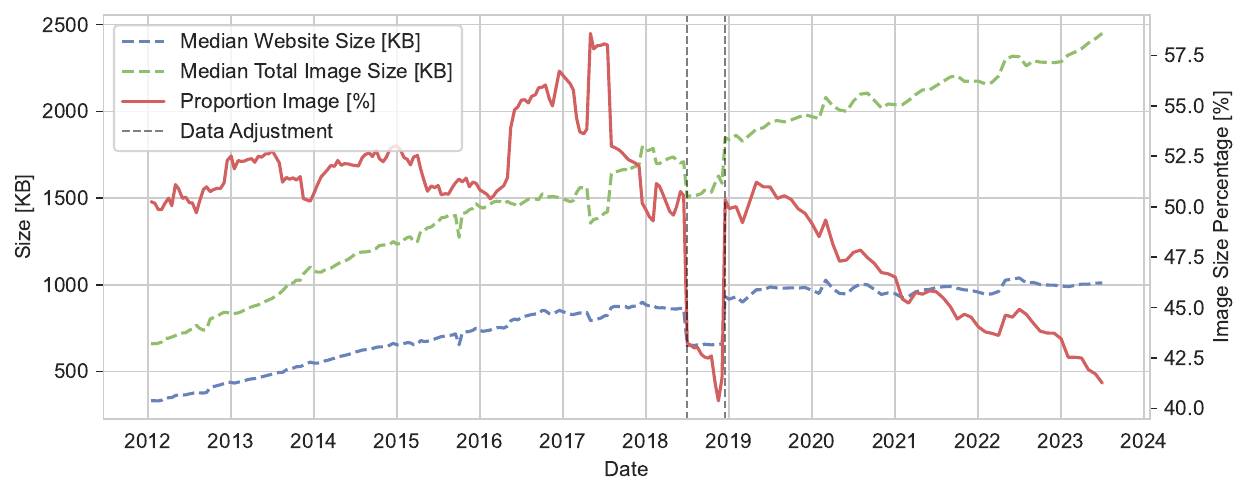}
    \caption{Total transfer size in kilobytes for all website resources, as well as separately requested images \cite{httparchive}.}
    \label{fig:sizeImageAndWebsite}
\end{figure}

These positive effects come with the drawback that often more data must be transmitted, which leads to higher energy demand \cite{morleyDigitalisationEnergyData2018}. As shown in Fig. \ref{fig:sizeImageAndWebsite}, an analysis of the data traffic shows that the median request size of the total website has increased over the decade, reaching 2449.4 KB in July 2023. Similarly, the total image file size has risen to 1010.6 KB in July 2023, although the percentile has slightly declined in recent years. Nevertheless, images still account for nearly 41\% of the total website request size, making it a big contributor to the environmental footprint of webpages \cite{jonesHowStopData2018}. This issue is caused not only by data transfer but also by image rendering. According to Thiagarajan et al. \cite{jonesHowStopData2018}, the "amount of energy used to render images is proportional to the number and size of images on the page." 

For web apps, various image formats can be used, such as \ac{PNG}, \ac{JPEG}, \ac{SVG}, or \ac{GIF}, which are well-established and widely supported. However, newer image formats, such as \ac{WEBP} and \ac{AVIF}, have emerged as alternatives that could offer better compression and faster loading times than older formats, leading overall to better performance. Due to length limitations, we refer the reader to \cite{frainResponsiveWebDesign2022} for a detailed explanation of the image formats. 
Until recently, these newer image formats often lacked browser support, as shown in Table \ref{tab:browserSupport}, and might not often be used as expected. 

\begin{table}[ht!]
\centering
\caption{First support year of image format in different browsers}
\label{tab:browserSupport}
\begin{tabular}{lR{1.8cm}R{1.8cm}R{1.8cm}R{2.8cm}R{1.8cm}}
\hline 
\textbf{Browser} & \textbf{Chrome} & \textbf{Safari} & \textbf{Firefox} & \textbf{Edge} & \textbf{Opera} \\
\hline 
\textbf{\ac{WEBP}} & 2014 (v32) & 2022 (v16) & 2019 (v65) & 2018 (v18) & 2014 (v19) \\
\textbf{\ac{AVIF}} & 2020 (v20) &  2023 (v16.4) & 2021 (v93) & Support (not off.) & 2020 (v71) \\
\hline 
\end{tabular}
\end{table}

The primary objective of this study is to compile and present an overview of the current utilization of images on the World Wide Web. To accomplish this, we applied web scraping techniques to gather data from the top 100,000 popular websites as of July 2023. Leveraging the collected website characteristics, we conducted a performance analysis focusing on raster-based image formats: PNG, JPEG, WEBP, and AVIF. This analysis underscores the potential performance improvements associated with these formats, considering a wide range of browsers. The specific contributions of this study are as follows:
\begin{enumerate}
    \item Investigation of prevalent image format preferences within web applications, accompanied by an analysis of their distribution and intrinsic characteristics across frequently-visited websites.
    \item Systematic assessment of the impact of various web image formats on application performance across a range of popular web browsers.
\end{enumerate}
After providing the necessary context and motivation for this research, the succeeding sections of this paper are organised
as follows. Subsequently, we discuss in Section \ref{sec:relatedWork} the findings of related secondary studies and emphasise the differentiation from our work. In Section \ref{sec:experimentalMethodology}, we outline the methodology employed in conducting Web Scraping and Web Performance Measurement, including the specific research questions. These questions are subsequently addressed in Section \ref{sec:results}, where we present the findings. The implications of these findings are then discussed in Section \ref{sec:discussion}. In Section \ref{sec:threats}, we address the potential threats to validity that may have influenced the results of the review and conclude with Section \ref{sec:conclusion}.

\section{Related Work}
\label{sec:relatedWork}
In order to streamline the incorporation of images into web applications, Zheng et al.~\cite{zhengResearchApplicationComputer2020} discuss the role of \textit{computer image processing} for manipulating and optimizing images to meet specific requirements. Specifically, they highlight common computer image processing steps to address web images. These steps encompass controlling image size, manipulating image shapes, adopting colors as well as transforming those into specific image formats. On the one hand, this can be done in a way that leads to optimal user satisfaction as well as can impact the performance of web applications.  

One specific study was done by Thiagarajan et al.~\cite{thiagarajanWhoKilledMy2012}, interested in the energy consumption of websites of mobile browsers. In the study conducted, the researchers have shown that rendering JPEG images is considerably cheaper and more energy-efficient than other formats, i.e., \ac{GIF} and \ac{PNG}, for equivalent-size 1600x1200 images. They claim that \ac{GIF}s were mainly used for small images like arrows and icons, while \ac{PNG}s were used for larger images like banners and logos. Moreover, \ac{JPEG}
images were used for handling even bigger images. The last mentioned, \ac{JPEG} seems to outperform the other formats due to its efficient encoding.

In 2021, Öztürk and Altan ~\cite{ozturkPerformanceEvaluationJPEG2021} examined the encoding process in more detail considering the lossless compression performance of the algorithms developed by the JPEG group (\ac{JPEG}-LS, \ac{JPEG} 2000, \ac{JPEG} XR, \ac{JPEG} XT, and \ac{JPEG} XL), \ac{WEBP}, as well as \ac{PNG}. Regardless of the image size, \ac{JPEG} XL exhibited the best compression ratios, along with satisfactory compression and decompression speeds. However, when it comes to decompression time, done on the client side, PNG images take less time compared to \ac{WEBP} and \ac{PNG}. 

Focusing on low-cost mobile hardware, Singh and Zaki \cite{singhComparativeEvaluationNextGeneration2023} examined in their report 1300 web pages (e.g., page size, economic context of the country of origin, and web page popularity). Furthermore, they have examed the performance impact of the "newly" image formats \ac{WEBP} and \ac{AVIF} in order to show the positive effect for low-cost hardware devices. They come to the conclusion that switching to these formats offers a simple optimization that significantly enhances web page agility and competitiveness in low-income countries.

Expanding on the aforementioned related work, we have incorporated web scraping into our methodology to understand the current image format landscape comprehensively. Building upon the insights from previous studies, our experiments were conducted across all commonly used browsers on two distinct client devices, offering a more holistic perspective.

\section{Experimental Methodology}
\label{sec:experimentalMethodology}
\begin{figure}[!ht]
    \centering
    \includegraphics[width=0.80\textwidth]{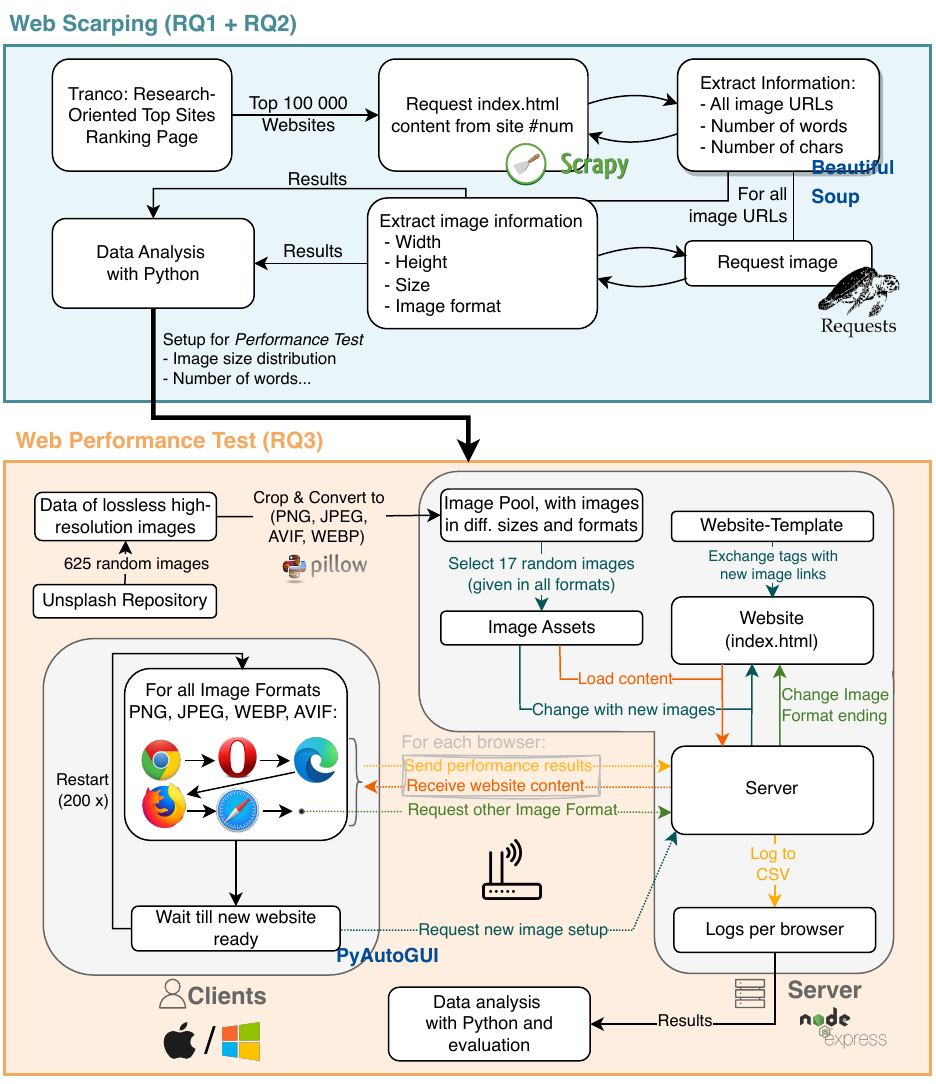}
    \caption{Overview of methodology and its techniques.}
    \label{fig:methodolgy}
\end{figure}

In this research, we conducted an exploratory study to examine how images are currently used in web applications and analyzed the hardware-related performance of various image formats on the web. To accomplish this, we formulated three research questions, used web scraping techniques and data analysis, and performed automated performance tests. To better illustrate the overall strategy, Figure \ref{fig:methodolgy} provides an overview of the general approach as well as shows the specific steps involved, described in the following subsections.

\subsection{Research Questions}
\label{sec:researchQuestions}
The objective of the research questions are to understand the usage of images on the web (\hyperref[sec:rq1]{RQ1}, \hyperref[sec:rq2]{RQ2}) and analyze the impact of specific image formats on the performance on the web (\hyperref[sec:rq3]{RQ3}).\\
\label{sec:rq1}\textbf{\textcolor{webScarping}{RQ1:} What are the prevailing image formats utilized across the most popular website on the World Wide Web?}\\
Over the last few years, several image formats have been common that are pixel-based with quite long-existing formats \ac{PNG} and \ac{JPEG}, "newer" \ac{AVIF} and \ac{WEBP} or the vector-based format \ac{SVG}. This research question aims to investigate the prevalence of various formats among a data set comprising \num{100 000} websites, already used by previous research. Specifically, we seek to determine whether \ac{WEBP}, despite being in existence for 13 years \cite{ImpactJPEGWebPConversion2016}, is not that popular compared to other formats.\\
\label{sec:rq2}\textbf{\textcolor{webScarping}{RQ2:} What are the predominant image characteristics found on popular websites, and how do these characteristics vary across different image formats?}\\
Building upon the results obtained from the analysis presented in \hyperref[sec:rq1]{RQ1}, this research question seeks to delve deeper into the discourse surrounding additional image characteristics. The attributes considered for investigation within this section include the following:
 \begin{itemize}
     \item \ac{BPP} per Image Format
     \item Image File Size (Width \& Height) Distribution
     \item Textual Findings
 \end{itemize}
\label{sec:rq3}\textbf{\textcolor{performanceTest}{RQ3:} What is the comparative impact of different web image formats and common web browsers on performance improvement in web applications?}\\
This research question is dedicated to investigating and contrasting the effects of raster-image formats, consisting of the established options  \ac{PNG} and \ac{JPEG}, as well as newer alternatives \ac{WEBP} and \ac{AVIF}. Non-raster formats like infinitely scalable SVGs and (multi-image) GIFs are excluded from consideration due to their structural disparities from raster-based image formats. The central objective is to examine the difference between these formats and prominent web browsers (i.e., Chrome, Edge, Safari, Opera, and Firefox) and their consequential influence on the loading speed of web content. 
    
\subsection{Web Scraping for Image Information}
\label{sec:webScarping}
In order to address the research questions designated as \hyperref[sec:rq1]{RQ1} and \hyperref[sec:rq2]{RQ2}, we have commenced the utilization of Web Scraping, a technique also identified as web extraction or harvesting. This approach enables the systematic retrieval of relevant data from diverse online sources \cite{zhaoWebScraping2017}. To automate the process \textit{Scrapy}\footnote{\url{https://scrapy.org}}, an open-source and collaborative Python framework is used. To feed Scrapy with a list of relevant \num{100000} websites, we used the \textit{Tranco} data set \cite{lepochatTrancoResearchOrientedTop2019}, claiming to be a research-oriented top site ranking, already used by several other studies. Websites, including limiting access by \texttt{robots.txt} or without any images included, were excluded. 
In the initial phase, we scraped through the \texttt{index.html} of each website, analyzing relevant data from the raw HTML code. Using \textit{Beautiful Soup}\footnote{\url{https://www.crummy.com/software/BeautifulSoup/}}, we extracted all image URLs and also gathered information about the text (i.e., number of words and characters) visible for users. For all extracted image URLs, we started a second run, extracting relevant information about the images used. Therefore, we used \textit{Request} to download the images and extract the necessary information (for details, see replication package \cite{dornauerWebImageFormats2023}). 
\subsection{Web Performance Measurements}
The second stage of the methodology involves a performance evaluation to investigate the effectiveness of different image formats in conjunction with a range of web browsers. Initially, we created a website with a simple design that avoids unnecessary features. This ensures any performance issues are primarily due to differences in web browsers and image formats. Leveraging the insights garnered from web scraping of real-world sites (see Section \ref{sec:ResultswebScarping}), we created a website with structured content that comprises 792 words and the integration of 17 image tags \texttt{<!-- Image \#nb -->}
which were substituted with their corresponding images. The rationale for choosing the image distribution is based on the following assumptions: 
\begin{enumerate}
    \item Image aspect ratios of 17 images:
    \begin{itemize}
        \item Landscape: 57\% (9 images)
        \item Symmetric: 32\% (6 images)
        \item Portrait: 11\% (2 images)
    \end{itemize}
    \item Common image dimensions (e.g., $1280\times720$, $64\times64$, ...), see Tab. \ref{fig:correlationWidthHeightDensity}
    \item One prominent large image \cite{portis2022WebAlmanac2022}
    \item One invisible tracking $1\times1$ pixel 
\end{enumerate}

Based on these assumptions, we derived a set of image formats, given in Table \ref{tab:selectedImageSize}. 
This collection of images counts a total of \num{1732621} pixels, which closely matches the median pixel count found on the scraped webpages (\num{1735713}).

In our image sampling, we randomly picked 625 high-resolution lossless images from Unsplash\footnote{\url{https://unsplash.com}}. The data set contained images that covered a wide spectrum of visual content. The photos were then trimmed to the different image dimensions and converted to the four image formats \ac{PNG}, \ac{JPEG}, \ac{WEBP}, and \ac{AVIF}. For this purpose, we selected the Python Imaging Library \textit{Pillow}\footnote{\url{https://pillow.readthedocs.io/}}, with over 11K GitHub stars, and also used by other scientific papers such as \cite{dhanawadeOpenCVBased2020}, \cite{huComparativePerformanceAnalysis2017}. To ensure consistent image quality, we adjusted the compression levels of each image format accordingly. Therefore, we used \ac{SSIM} equal to 0.95 (compared to PNG) as the criterion, as this implies almost perfect similarity \cite{horeImageQualityMetrics2010,setiadiPSNRVsSSIM2021} among the other image formats.

\begin{table}[ht!]
\centering
\caption{Selection of image sizes in pixels, based on \hyperref[sec:rq2]{RQ2} }
\label{tab:selectedImageSize}
\begin{tabular}{R{4.5cm}R{4.5cm}R{2.5cm}} 
\hline
\rowcolor[rgb]{0.753,0.753,0.753} \multicolumn{1}{c}{Landscape} & \multicolumn{1}{c}{Symmetric} & \multicolumn{1}{c}{Portrait} \\ 
\hline
3 x (240, 180), 2 x (\phantom{0}300, 225) & 1 x (\phantom{00}1, \phantom{00}1), 1 x (\phantom{0}32, \phantom{0}32) & 1 x (225, 300) \\
2 x (320, 180), 1 x (\phantom{0}640, 360) & 1 x (\phantom{0}64, \phantom{0}64), 1 x (120, 120)                     & 1 x (300, 420) \\
1 x (1280, 720)                 & 1 x (150, 150), 1 x (200, 200)                                          &  \\
\hline
\end{tabular}
\end{table}

The test automation setup consisted of a server (Express  4.18.2), a router (Netgear WNDR 4300), and two client machines (1) MacBook Pro 2019 16 Zoll (2.3 GHz 8‑Core Intel Core i9, 16 GB) with MacOS 13.5 as well as (2) ThinkPad T490 (1,80 GHz, Intel Core i7, 16 GB) with Windows 11. 

Besides hosting the website, the web server additionally had the ability to randomly choose images (using the same configuration) and alternate between their respective image formats. Furthermore, it collected measurements from the client machines. These machines were connected via WAN in an isolated environment, i.e. with no access to the internet or other clients.  

The setup for each client machine is based on the default settings provided by the operating system supplier, excluding any third-party software. We proceeded to install a selection of the most up-to-date and widely used web browsers on each machine, guided by the browser usage data from July 2023 as presented by \textit{StatCounter}\cite{BrowserMarketShare}: Chrome ($63.33\%$), Safari ($19.95\%$), Edge ($5.14\%$), Opera ($2.98\%$), Firefox ($2.79\%$) and Others ($5.81\%$). It's important to note that Safari was exclusively employed on MacOS due to the specific OS prerequisites. Additionally, Python 3.11 was installed on all machines, along with the \textit{PyAutoGUI}\footnote{\url{https://pyautogui.readthedocs.io}} and \textit{Keyboard}\footnote{\url{https://pypi.org/project/keyboard/}} libraries, enabling us to automate the execution of the tests. 

An automated test run consisted of three steps for each browser. First, the Incognito/Private Browser was launched to eliminate the influence of cookies and caching. Second, the URL was entered, and the webpage was loaded while its performance was measured. Third, the browser was closed, terminating the session and initiating a cooldown period before the next run. A test cycle comprised 20 (for Windows: 16) experimental iterations, each employing a distinct set of images randomly selected from a pool. The test cycle systematically traversed 5 (for Windows: 4) different browsers in sequence and then transitioned to the subsequent four image formats. We focused on the metrics
\begin{itemize}
    \item \textbf{\ac{PLT}}:  time from the start of page loading to the display of any part of the page’s content on the screen, and
    \item \textbf{\ac{FCP}}: which measures the time from the start of page loading to the end of page rendering 
\end{itemize}
to measure the performance of each test run. These metrics reflect how well a web application performs, as shown by \cite{departmentofcomputerscienceengineeringstamforduniversitybangladeshbangladeshWebApplicationPerformance2021, pourghassemiWhatIfAnalysisPage2019}. However, other metrics are also available but not considered for this evaluation. For these measurements, we used the \textit{PerfumeJS}\footnote{\url{https://zizzamia.github.io/perfume/}} performance profiler with the latest stable release v8.4.0. It leverages the latest Performance APIs to collect field data for cross-browser testing and was used in previous publications \cite{malavoltaJavaScriptDeadCode2023, malavoltaFrameworkAutomaticExecution2020}. 

The experimental design employed in this study follows a 4-5 (for Windows: 4-4) Factorial Design principle, wherein the combinations of two independent variables, namely browser types and image formats, are systematically manipulated to investigate their joint impact on web page loading performance. By utilizing the \textit{Scheirer–Ray–Hare test}\cite{scheirerAnalysisRankedData1976}, a non-parametric version of two-way ANOVA based on ranks, on this factorial arrangement, we gain an understanding of how each variable independently and collectively contributes to the observed variations in the measured metrics.

\section{Results}
To address the aforementioned research questions, we have employed web scraping techniques to gather relevant data. The results obtained from the web scraping analysis are presented in the following subsection \ref{sec:ResultswebScarping} (\hyperref[sec:rq1]{RQ1} and \hyperref[sec:rq2]{RQ2}). Moreover, we address \hyperref[sec:rq3]{RQ3} in subsection \ref{sec:resultsPerformanceTest}, presenting the results of the web performance test.  

\label{sec:results}
\subsection{Results using Web Scraping}
\label{sec:ResultswebScarping}
We scraped the \num{100 000} most frequently visited websites listed in Tranco \cite{lepochatTrancoResearchOrientedTop2019}, which focuses on research-oriented top-site ranking. After filtering out websites that were blocked by \texttt{robot.txt} or have not contained any images, we were able to analyze \num{58057} sites. This resulted in a total of \num{2662548} images in 10 formats, namely \ac{JPEG}, \ac{PNG}, \ac{SVG}, \ac{GIF}, \ac{WEBP}, \ac{BMP}, \ac{ICO}, \ac{MPO}, \ac{PSD}, and \ac{TIFF} formats, with a median of 25 images per site (17 unique images - duplicates are images with same URL). 

\subsubsection{Image Format Distribution:}
\begin{figure}
    \centering   
    \includegraphics[width=0.9\textwidth]{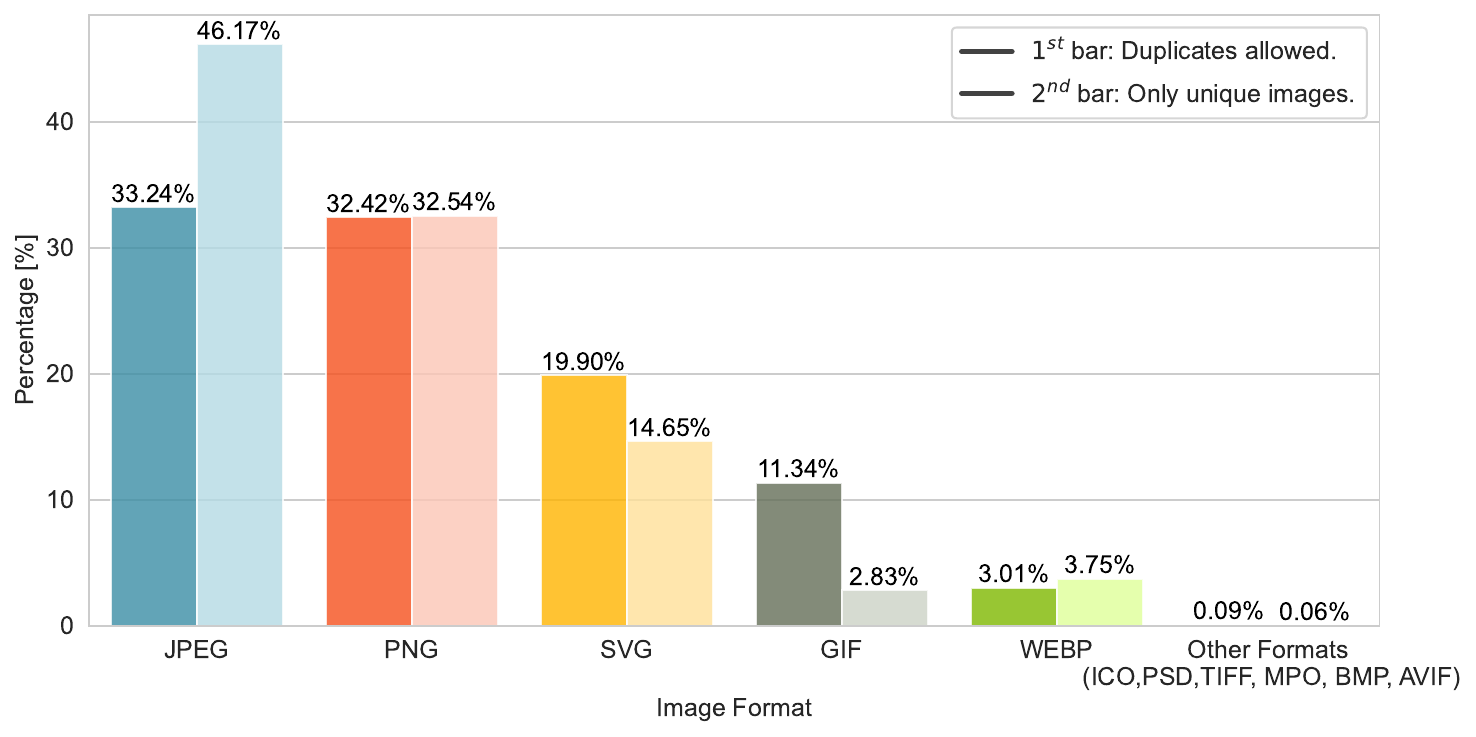}
    \caption{Distribution plot of image formats among the \num{2662548} images collected from crawled websites.}
    \label{fig:distributionFigureFormats}
\end{figure}

Considering the different formats, we calculated the distribution of the images regarding their image format, illustrated in Fig. \ref{fig:distributionFigureFormats}. Thereby, we found that \ac{JPEG} and \ac{PNG} formats are the most commonly used, with each format making up almost one-third of all images. SVG comes in third place at 19.90\%, followed by GIF at 11.34\%. It's worth noting that \ac{WEBP} format only makes up 3.01\% of occurrences, indicating it is not as commonly used. All other formats \ac{BMP}, \ac{ICO}, \ac{MPO}, \ac{PSD}, \ac{BMP}, \ac{AVIF} and \ac{TIFF} were not that popular by web developers, in total 0.05\% of all images. An interesting observation to consider is that some images occur multiple times. This specifically includes the formats \ac{SVG} and \ac{GIF}. 

\subsubsection{Bits per Pixel:}
\begin{figure}
    \centering  
    \includegraphics[width=0.9\textwidth]{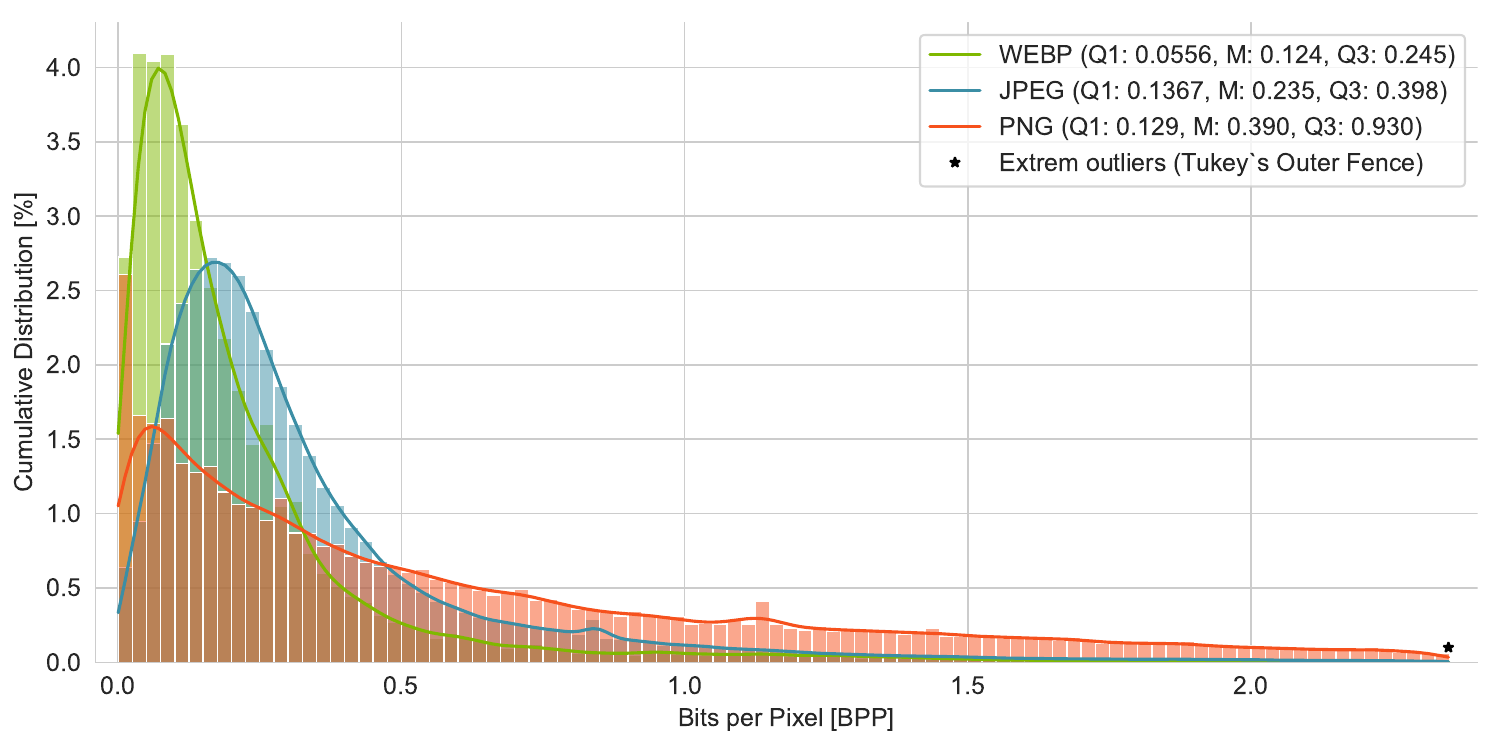}
    \caption{Distribution of the \ac{BPP} for the image formats \ac{JPEG}, \ac{PNG}, and \ac{WEBP}. The extreme outliers are not illustrated, based on Tukey's Outer Fence \cite{tukeyExploratoryDataAnalysis1977} and checked to justify their exclusion. The related quartiles 1,3 (Q1, Q3) and median (M) are given as (Q1, M, Q3)}.
    \label{fig:bitsPerPixelDistribution}
\end{figure}

We have depicted the distribution of \ac{BPP} for the image formats JPEG, PNG, and WEBP, as shown in Figure \ref{fig:bitsPerPixelDistribution}. The distribution is presented using discrete bins with an interval of 0.025 \ac{BPP}, along with the Kernel Density Estimate (KDE). Within the \ac{BPP} range of $[0.0, 0.25]$, it is observed that WEBP exhibits the highest frequency of occurrences, suggesting that a larger proportion of images, as compared to JPEG and PNG, possess a smaller file size per pixel. In general, the majority of PNG images utilize the highest file size per pixel when compared to the other image formats studied in this analysis. 

\subsubsection{Width and Height Properties:}
\begin{figure}[ht!]
\centering
    \begin{subfigure}{0.49\textwidth}
        \includegraphics[width=1\textwidth]{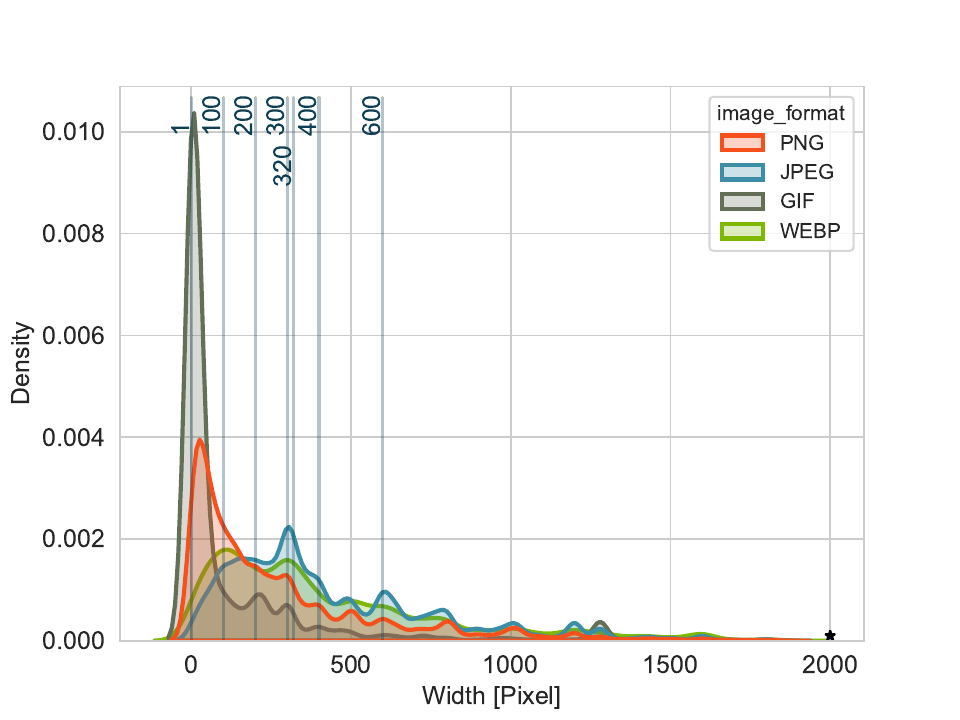}
        \caption{Density of \textbf{Width} per image format.}
        \label{fig:widthDensity}
    \end{subfigure}
\hfill
    \begin{subfigure}{0.49\textwidth}
        \includegraphics[width=1\textwidth]{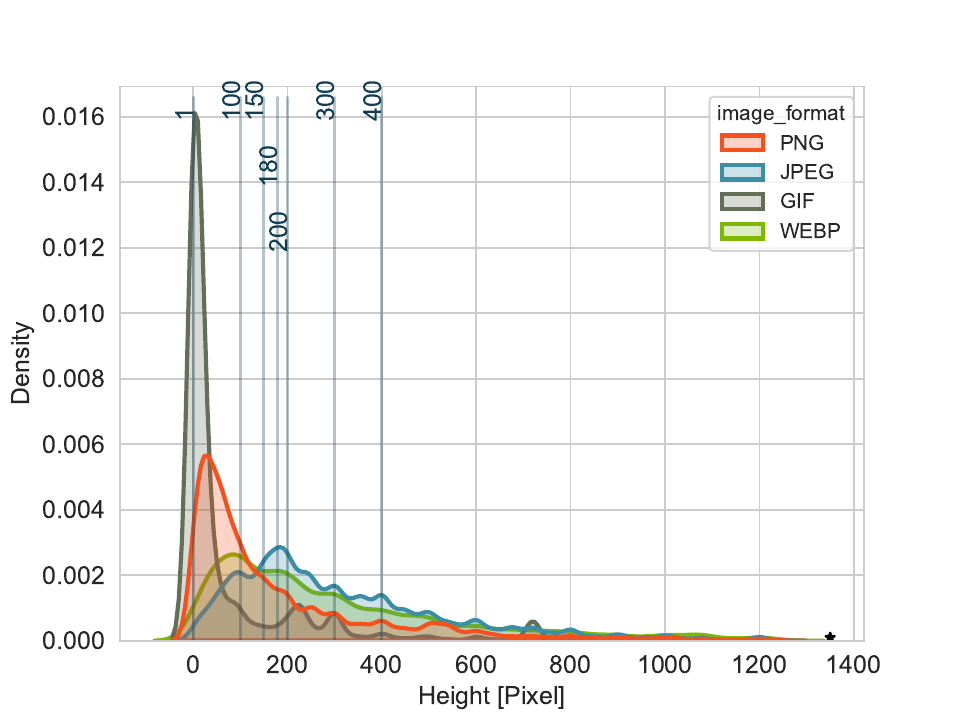}
        \caption{Density of \textbf{Height} per image format.}
        \label{fig:heightDensity}
    \end{subfigure}
\hfill
    \begin{subfigure}{0.7\textwidth}
        \includegraphics[width=\textwidth]{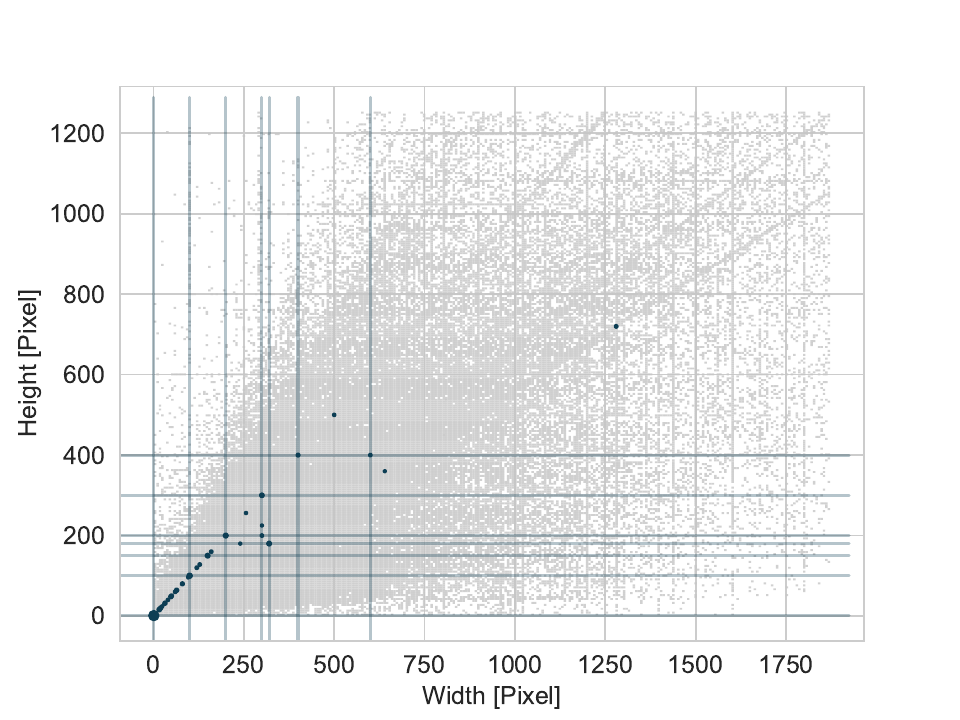}
        \caption{Scatter plot, using width and height. The dark-blue \\dots indicate the 30 most frequent image sizes.}
        \label{fig:correlationWidthHeightDensity}
    \end{subfigure}
    \begin{subfigure}{0.29\textwidth}
        \centering
       \begin{tabular}{rrr}
            \cline{2-3}
            \multicolumn{1}{c}{} & \multicolumn{2}{c}{\cellcolor[HTML]{C0C0C0}Density} \\ \hline
            \rowcolor[HTML]{C0C0C0} 
            \multicolumn{1}{c}{Pixel} & \multicolumn{1}{c}{\cellcolor[HTML]{C0C0C0}Width} & \multicolumn{1}{c}{\cellcolor[HTML]{C0C0C0}Height} \\ \hline
            $[0,100]$& \multicolumn{1}{r}{$ 28.3\%$} & $36.8\%$ \\
            $(100,200]$ & \multicolumn{1}{r}{$ 15.9\%$} & $20.5\%$ \\
            $(200,300]$ & \multicolumn{1}{r}{$ 14.9\%$} & $14.0\%$ \\
            $(300,400]$ & \multicolumn{1}{r}{$ 11.2\%$} & $9.1\%$ \\
            $(400,500]$ & \multicolumn{1}{r}{$ 6.1\%$} & $6.3\%$ \\
            $(500,600]$ & \multicolumn{1}{r}{$ 5.4\%$} & $4.8\%$ \\
            $(600,700]$ & \multicolumn{1}{r}{$ 3.8\%$} & $2.5\%$ \\
            $(700,800]$ & \multicolumn{1}{r}{$ 4.3\%$} & $2.5\%$ \\
            $(800,900]$ & \multicolumn{1}{r}{$ 1.9\%$} & $1.2\%$ \\
            $(900,1000]$ & \multicolumn{1}{r}{$ 1.9\%$} & $1.0\%$ \\
            $(1000,1100]$ & \multicolumn{1}{r}{$ 1.5\%$} & $0.7\%$\\
            $(1100,1200]$ & \multicolumn{1}{r}{$ 1.7\%$} & $0.5\%$\\
            $(1200,\infty)$ & \multicolumn{1}{r}{$ 3.1\%$} & $0.1\%$\\
            \hline 
        \end{tabular}
        \caption{Density distribution per 100 chunks.}
        \label{tab:densityTable}
    \end{subfigure}
\caption{The plots depict width density (Figure \ref{fig:widthDensity}), height density (Figure \ref{fig:heightDensity}), and their correlation (Figure \ref{fig:correlationWidthHeightDensity}). Extreme outliers are excluded using Tukey's Outer Fence \textbf{($\ast$)}. Standard normalization is applied to each font format for Figures \ref{fig:widthDensity} and \ref{fig:heightDensity}.}
\label{fig:widthHeightExplaination}
\end{figure}

Due to their resolution independence and infinite scalability, SVG images were excluded from the analysis, focusing solely on raster images (PNG, JPEG, GIF, and WebP) to assess width and height. The units employed in our analysis are denoted in pixels.

As an initial point, we removed the extreme outliers using Tukey's Outer Fence equally for the width and height of the images, which led to the inclusion of $97.31\%$ of all images. Using the data, we started to illustrate the density of the frequency of specific widths (Fig. \ref{fig:heightDensity}) in the range of $[0, 1872]$ as well as the density of heights (Fig. \ref{fig:widthDensity}) in the range of $[0,1254]$ for the image formats independently. 

Overall, we discern that nearly all image dimensions are well-represented. Additionally, conspicuous patterns emerge, highlighting certain values that appear with higher frequencies. Specifically, for width, the prominent values are $1$, $300$, $200$, $100$, $320$, $600$, while for height, they are $1$, $200$, $300$, $100$, $180$, $150$, $400$. These values are presented in descending order of their prevalence. Notably, we observed a prevalence of sizes multiples of a hundred.

Notably, the top ten dimensions comprise $1 \times 1$ (Invisible Pixels (tracking) \cite{bekosHitchhikerGuideFacebook2023}), $100 \times 100$, $150 \times 150$, $320 \times 180$, $16 \times 16$, $320 \times 180$, and other symmetric dimensions such as ($16$, $300$, $64$, $32$). 

It appears that widths are more frequently longer ($Q1: 96px$, $M: 262px$, $Q3:540px$ ) compared to heights ($Q1: 64px$, $M: 180px$, $Q3: 380px$), as shown by Tab.\ref{tab:densityTable}. Our observations reveal that $57\%$ of the images exhibit a Landscape orientation, wherein their width exceeds their height horizontally. In contrast, $31.22\%$ of the images possess a symmetric aspect ratio, while a smaller proportion, constituting only $11.78\%$, are in Portrait mode, where their height surpasses their width vertically (ref. Fig. \ref{fig:widthHeightExplaination}).

\subsubsection{Textual Findings:}
In addition to conducting image-related analyses, we also compiled data regarding the textual content present on the examined websites. The median word count per site was determined to be $792$ words, with $Q1=407$ and $Q3=1397$.

\subsection{Results of the Web Performance Tests}
\label{sec:resultsPerformanceTest}
To evaluate the performance, we conducted 200 test runs on two different client devices and were able to measure the \ac{FCP} and \ac{PLT} metrics using PerfumeJS, supported by all browsers. 

\subsubsection{First Contentful Paint: } The \ac{FCP} metric measures the time from the start of page loading to the display of any part of the page’s content on the screen. We found no significant effect of the image formats on this metric, as the p-value was greater than 0.05. However, we observed a significant effect of the browsers on the \ac{FCP} metric, indicating that different browsers render the page content at different speeds.

\begin{figure}
    \centering
    \includegraphics[width=0.9\textwidth]{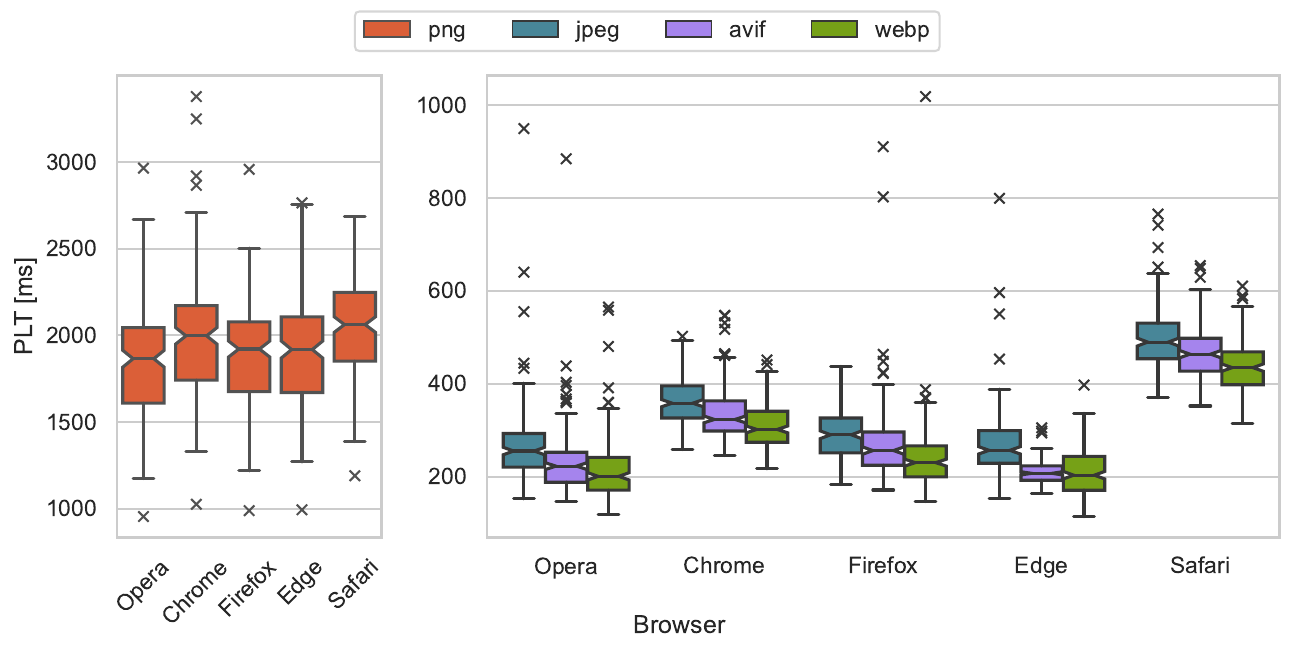}
    \caption{\ac{PLT} for different browsers and image formats on the Mac client. The results for the Windows client draw a similar picture.}
    \label{fig:PLTperBrowserAndFormat}
\end{figure}

\subsubsection{Page Load Time: } Considering the \ac{PLT} metric, which measures the time from the start of page loading to the end of page rendering (including loading and displaying all the content), we found that both the browser and the image formats had a significant effect on this metric, based on the results of the \textit{Scheirer–Ray–Hare test}\cite{scheirerAnalysisRankedData1976}, a non-parametric version of two-way ANOVA. The \ac{PNG} format had the highest median \ac{PLT} of 1943 ms, which was much longer than the other formats: \ac{JPEG} (312 ms), \ac{AVIF} (271 ms), and \ac{WEBP} (258 ms) for all browsers on the Mac. Figure \ref{fig:PLTperBrowserAndFormat} shows the breakdown of \ac{PLT} for each browser and image format. From this figure, we can notice that the performance gains from using different formats are consistent across different browsers. However, the browser itself also has a large influence on the \ac{PLT}, with Safari being the slowest and Opera being the fastest. The results for the Windows client show a nearly similar picture, demonstrating the independence of results from the client device.  

\section{Discussion}
\label{sec:discussion}
Images have a significant impact on human perception as they capture user attention more effectively than text \cite{wangVisualDesignWeb2021}. By using web scraping techniques (\textcolor{webScarping}{RQ1} \& \textcolor{webScarping}{RQ2}) , we have observed that the median number of images on websites was 25 (including header images, logos, and tracking pixels), showing their relevance for web apps. 

To dispaly raster-based images, two formats are commonly utilized: \ac{PNG} for smaller images and \ac{JPEG} for larger images. These findings are in line with research from previous years conducted by HTTP Archive \cite{httparchive}. However, we found that \ac{WEBP} and \ac{AVIF} are still uncommon in July 2023. This situation might arise due to the fact that support for \ac{WEBP} was lacking for certain browsers in the last years, specifically in the case of Safari. The picture for \ac{AVIF} is even worse (see Table \ref{tab:browserSupport}). 

However, images constitute a large fraction of the data transfer — up to 41\% — on the web, and the novel formats \ac{WEBP} and \ac{AVIF} offer potential benefits in reducing data consumption. Singh \cite{singhComparativeEvaluationNextGeneration2023} showed that replacing images on real-world websites with \ac{WEBP} and \ac{AVIF} resulted in a median decrease of 25\% (websites high-income countries) and 35 \% (websites low-income countries), in terms of page size. Considering the findings of our performance test (\textcolor{performanceTest}{RQ3}), we similarly identify a clear performance improvement, which comprises the transmitting, rendering, and displaying of content in a benchmark setup. The \ac{PLT} improves by a factor of 1.21 using \ac{WEBP} and by 1.15 using \ac{AVIF} compared to the \ac{JPEG} format. These findings are relevant not only for Chromium-based browsers (Chrome, Edge, Opera) but also for Safari and Firefox. 

Considering the situation with image format support and how they affect how fast websites load, we strongly suggest using the \ac{WEBP} format. \ac{WEBP} works well across the board, leading to increased website performance and a better user experience.

Besides that, newer image formats also extend the battery life of devices. This may seem insignificant, but it has a significant impact on energy conservation \cite{morleyDigitalisationEnergyData2018} considering a worldwide perspective. For example, if we  assume savings of 5\% through the new image formats for mobile devices, we would be able to shut down one of the Fukushima nuclear power plants \cite{dornauerEnergySavingStrategiesMobile2023}.

\section{Threats to Validity}
\label{sec:threats}
Threats to Validity are potential sources of error or bias that may compromise the quality and applicability of a study. Following Cook and Campbell \cite{cookQuasiexperimentationDesignAnalysis1979}, we organised our threats into the four types: \\
\textbf{Internal Validity: } To mitigate this risk, we conducted our experiment in a controlled and isolated environment, with consistent hardware and software configurations throughout all experimental runs. In addition, we automated the test execution using PyAutoGUI, allowing independent replication. Performance measurement stability was ensured through timed intervals, specific configurations to minimize external effects (e.g., disabled cache or turning off third-party software), and 200 test runs per image combination for outlier detection. One important threat is the influence of image attributes like color depth, distortion, and sharpness \cite{inouriFastEfficientApproach2018} on the image format efficiency. Thus, we employed a randomized selection strategy. From a pool of images, we randomly selected 17 images of various sizes, representing typical website diversity, based on results from \hyperref[sec:rq2]{RQ2} for each test run. Furthermore, we fall back to the \ac{SSIM} metric to ensure comparable image quality across all formats. \\
\textbf{External Validity: } The study examined a relatively small fraction of the existing websites, which may limit the generalizability of the results. However, the sample size of \num{58 057} websites, selected from the 100K most popular sites based on a representative dataset Tranco \cite{lepochatTrancoResearchOrientedTop2019}, should mitigate this threat. A comparison with gray literature \cite{portis2022WebAlmanac2022} showed a consistent pattern of the findings. Regarding the comprehensive conclusion drawn that \ac{WEBP} and \ac{AVIF} are recommended today for use in web applications, we conducted our study on two different operating systems and the five most popular web browsers (latest versions). We believe exploring the potential combinations of image formats could be another aspect worth investigating in more detail.\\ 
\textbf{Construct Validity: } The study's approach involves webpage design to isolate browser and format-related performance issues. An image dataset is utilized, and images are randomly selected, ensuring a wide range of visual content representation. The analysis employs a factorial design to dissect the interplay between browser types and image formats on the loading performance, employing common metrics for evaluation such as \ac{FCP} or \ac{PLT}. A detailed description is provided in the replication package on Zenodo \cite{dornauerWebImageFormats2023}. \\
\textbf{Statistical Conclusion Validity: } Employing a factorial design, we systematically manipulated both web browsers and image formats. This approach facilitated the isolation and thorough analysis of the distinct and combined impacts they exert on the loading performance of web pages. Consequently, the \textit{Scheirer–Ray–Hare test}\cite{scheirerAnalysisRankedData1976} was employed as the chosen statistical method for our analysis.

\section{Conclusion}
\label{sec:conclusion} 
Web images are pivotal for optimizing websites and enhancing user satisfaction by captivating their attention. However, they can prolong a website’s loading time due to bandwidth and rendering demands. Our analysis of popular websites highlights a predominant use of traditional image formats like \ac{PNG} and \ac{JPEG}. Based on performance tests across widespread browsers, we recommend adopting modern image formats, particularly \ac{WEBP} and \ac{AVIF}. Of these, \ac{WEBP} is preferable due to its broad compatibility and more efficient performance advantages. 

\section{Acknowledgment}
This project received support from ITEA3-SmartDelta and funding from the Austrian Research Promotion Agency (FFG, Grant No. 890417).
\bibliography{PROFES_Web_Images}
\end{document}